\documentclass[aps,prd,tightenlines,showpacs,nofootinbib,preprint]{revtex4}
\usepackage{colordvi}

\usepackage{amsbsy}

\usepackage{amsfonts}

\usepackage{amssymb}

\usepackage{amsmath}

\usepackage{graphicx}
\usepackage{epsfig}
\def\ba{\begin{eqnarray}}
\def\ea{\end{eqnarray}}
\def\be{\begin{equation}}
\def\ee{\end{equation}}
\def\tE{\tilde{E}}
\def\half{{\textstyle{1\over2}}}

\def\G{{\Gamma}}   \def\H{{\cal H}} 
\def\g{{\gamma}}  \def\a{{\alpha}} 
\def\U(1){{\rm U(1)}}   \def\S{{\cal S}}

\def\d{{\rm d}}

\def\bar{\overline}
\def\ii{{\rm i}}
\def\e{{\rm e}}

\newcommand{\teta}{\rlap{\lower2ex\hbox{$\,\tilde{}$}}\eta{}}

\def\={\mathrel{\widehat\mathalpha{=}}}
\def\puto#1{\rlap{\raise.5ex\hbox{\char'27}}{#1}}

\begin{document}
\preprint{\vbox{\baselineskip=12pt \rightline{ICN-UNAM-05/06}
\rightline{gr-qc/0507038} }}



\title{Loop Quantum Geometry: A primer}

\date{July 8th 2005}




\author{Alejandro Corichi}
\email{corichi@nucleares.unam.mx} \affiliation{Instituto de
Ciencias Nucleares, Universidad Nacional Aut\'onoma de M\'exico,
A. Postal 70-543, M\'exico D.F. 04510, M\'exico}




\begin{abstract}
This is the written version of a lecture given at the ``VI Mexican
School of Gravitation and Mathematical Physics" (Nov 21-27, 2004,
Playa del Carmen, M\'exico), introducing the basics of Loop
Quantum Geometry. The purpose of the written contribution is to
provide a Primer version, that is, a first entry into Loop Quantum
Gravity and to present at the same time a friendly guide to the
existing pedagogical literature on the subject. This account is
geared towards graduate students and non-experts interested in
learning the basics of the subject.
\end{abstract}

\pacs{04.60.Pp}
 \maketitle

\section{Introduction}

\noindent {\it Loop Quantum Gravity} (LQG) has become in the past
years a mayor player as a candidate for a quantum theory of
gravity. On the one hand it has matured into a serious contender
together with other approaches such as {\it String/M Theory}, but on the
other is it not as well understood, neither properly credited as a
real physical theory of quantum gravity. The purpose of this
contribution is to provide a starting point for those interested
in learning the basics of the theory and to provide at the same
time an introduction to the rich literature on the subject which
includes very well written reviews and monographs. Given the space
constraint we shall not attempt to write a comprehensive review of
LQG, but to provide, we hope, and useful guide to the subject.

Let us start by providing a list of references that will be useful
in the various stages. Firstly, there are several primer
introductions to the subject, written for different purposes. For
instance, there was for many years the canonical primer by Pullin
\cite{pullin}. Unfortunately, it is now somewhat dated. Good
introductions to spin networks and recoupling theory needed in LQG
are given by the primers by Rovelli \cite{Rovelli:1998gg} and
Major \cite{seth}. There are recent up-to-date accounts written
for non-experts that give nice motivation, historical perspective
and an account of recent and in progress work from two different
perspectives \cite{AA:NJP} and \cite{Smolin:2004sx}. There are
also technical reviews that give many details and are certainly a
good read \cite{AL:review}, \cite{Perez:2004hj},
\cite{Thiemann:2002nj}, and (from an outside perspective)
\cite{Nicolai:2005mc}.

The subject has matured enough so that several monographs have
been written, including some recent and updated. These monographs
approach and present the subject from  different perspectives
depending, of course, on the authors own taste. From these, it is
worth mentioning two. The first one by Rovelli is  physically
motivated but not so heavy in its  mathematical treatment, and can
be found in \cite{Rovelli:2004tv}. A mathematically precise
treatment, but not for the faint of heart is given by the
monograph by Thiemann \cite{Thiemann:2001yy}. There have been also
several nice reviews that motivate and give a birdseye view of the
subject such as \cite{Rovelli:1999hz}, \cite{Rovelli:1997yv} and
\cite{pullin2}. Finally, there are several accounts on comparisons
between loop quantum gravity and other approaches, such as string
theory. On chronological order, we have a review by Rovelli
\cite{Rovelli:1997qj}, an entertaining dialog
\cite{Rovelli:2003wd} and a critical assessment by Smolin
\cite{Smolin:2003rk}.

The second purpose of this paper is to present an introduction to
the formalism known as {\it(loop) quantum geometry}. The
difference between quantum geometry and loop quantum gravity is
that the former is  to be thought of as the (new) formalism
dealing with background-free quantum theories based on
connections, whereas the later is a particular implementation
where gravity, as defined by general relativity, is the theory
under consideration. For instance, one could think of applying the
same formalism to more general theories such as supergravity
and/or higher derivative theories. In the remainder of this
section we shall give a motivation for why one should study loop
quantum gravity, when one is interested in the basic problem of
uniting quantum mechanics and the theory of gravitation.

Why should one study loop quantum gravity? For one thing, it is
based on two basic principles, namely the general principles of
quantum theory and one of the main lessons from general
relativity: that physics is diffeomorphism invariant. This means
that the field describing the gravitational interaction, and the
geometry of spacetime is fully dynamical and interacting with the
rest of the fields present. When one is to consider its quantum
description, this better be background independent. The fact that
LQG is based in general principles of quantum mechanics means only
that one is looking for a description based on the standard
language of quantum mechanics: states are elements on a Hilbert
space (well defined, of course), observables will be Hermitian
operators thereon, etc. This does not mean that one should use
{\it all} that is already known about quantizing fields. Quite on
the contrary, the tools needed to construct a background
independent quantization (certainly not like the quantization we
know), are rather new.

Another reason for studying LQG is that this is the most serious
attempt to perform a full {\it non-perturbative} quantization of
the gravitational field. It is an attempt to answer the following
question: can we quantize the gravitational degrees of freedom
without considering matter on the first place? Since LQG aims at
being a physical theory, which means it better be falsifiable, one
expects to answer that question unambiguously, whenever one has
the theory fully developed. This is one of the main present
challenges of the theory, namely to produce predictions that can
be tested experimentally. Since the theory does not suffer from
extra dimensions nor extended symmetries, one expects that the
task will be feasible to complete, without the burden of getting
rid on those extra features. Will this be the final theory
describing the quantum degrees of freedom of the gravitational
field? Only experiments will tell, but for the time being, LQG
remains an intriguing possibility very well worth the trial.

This contribution is organized as follows: In Sec.~\ref{sec:2} we
provide some of material needed in order to be ready to fully
grasp the details of LQG. In particular, we recall the canonical
description go general relativity in the geometrodynamics language
and perform the change of variables to go to the description in
terms of connections. In Sec.~\ref{sec:3} we consider within the
classical description of the gravitational field, the observables
that will be regarded as the basic objects for the quantization.
We shall see that both background independence and diffeomorphism
invariance lead us to select holonomies and electric fluxes as the
basic objects. Section~\ref{sec:4} will be devoted to the
discussion of the Hilbert space of the theory, its multiply
characterizations and a particular basis that is very convenient,
namely, the spin network basis. In Section~\ref{sec:5} we provide
a discussion of the new (quantum) geometry that the theory
presents for us, focusing on basic geometrical operators.
Section~\ref{sec:6} will be devoted to a list of accomplishments
and open issues.

Finally,  we should note that our list of references is
minimalist, trying to concentrate on review articles or monographs
rather than in the original articles. A more complete reference
list can be found in the books \cite{Thiemann:2001yy} and
\cite{Rovelli:2004tv} and in the bibliography compilation
\cite{biblio}. An online guide for learning LQG with references
can be found in \cite{seth2}.

\section{Preliminaries}
\label{sec:2}


\noindent What are the pre-requisites for learning LQG? First of
all, a reasonable knowledge of General Relativity (GR), specially
its Hamiltonian formulation as in \cite{wald1} and \cite{poisson},
acquaintance with QFT on flat space-time and preferably some
notions of ``quantization", namely the passage from a classical
theory to a quantum one as in \cite{wald2}. Finally, the language
of gauge field theories is essential, including connections and
holonomies. A good introduction to the subject, including the
necessary geometry, is given in \cite{baez}.

The first step is to introduce the basic classical variables of
the theory. Since the theory is described by a Hamiltonian
formalism, this means that the 4-dim spacetime $M$ is of the form
$M=\Sigma\times\mathbb{R}$, where $\Sigma$ is a 3-dimensional
manifold. The first thing to do is to start with the
geometrodynamical phase space $\G_{\rm g}$ of Riemannian metrics
$q_{ab}$ an $\Sigma$ and their canonical momenta
$\tilde{\pi}^{ab}$ (related to the extrinsic curvature $K_{ab}$ of
$\Sigma$ into $M$  by
$\tilde{\pi}^{ab}=\sqrt{q}\,(K^{ab}-\frac{1}{2}q^{ab}\,K)$, with
$q={\rm det}(q_{ab})$ and $K=q^{ab}K_{ab}$). Recall that they
satisfy,
\be
\{\tilde{\pi}^{ab}(x),q_{cd}(y)\}=2\kappa\,\delta^{a}_{(c}\,\delta^b_{d)}
\,\delta^3(x,y)\quad
;\quad
\{q_{ab}(x),q_{cd}(y)\}=\{\tilde{\pi}^{ab}(x),\tilde{\pi}^{cd}(y)\}=0
\ee
General Relativity in these geometrodynamical variables is a
theory with constraints, which means that the canonical variables
$(q_{ab},\tilde{\pi}^{ab})$ do not take arbitrary values but must
satisfy four constraints: \be {\cal
H}^b=D_a\,(\tilde{\pi}^{ab})=\approx 0 \qquad {\rm and},\qquad
{\cal
H}=\sqrt{q}\,\left[R^{(3)}+q^{-1}(\half{\tilde{\pi}^2}-\tilde{\pi}^{ab}
\tilde{\pi}_{ab})\right]\approx 0 \ee The first set of constraints
are known as the vector constraint and what they generate (its
gauge orbit) are spatial diffeomorphisms on $\Sigma$. The other
constraint, the scalar constraint (or super-Hamiltonian) generates
``time reparametrizations". We start with 12 degrees of freedom,
minus 4 constraints means that the constraint surface has 8
dimensions (per point) minus the four gauge orbits generated by
the constraints giving the four phase space degrees of freedom,
which corresponds to the two polarizations of the gravitational
field.

In order to arrive at the connection formulation, we need first to enlarge
the phase space $\G_{\rm g}$ by considering not metrics $q_{ab}$ but the
 co-triads $e_a^i$ that define the metric by,
\be q_{ab}=e_a^i\,e_b^j\,\delta_{ij} \ee where $i,j=1,2,3$ are
internal labels for the frames. These represent 9 variables
instead of the 6 defining the metric $q_{ab}$, so we have
introduced more variables, but  at the same time a new symmetry in
the theory, namely the $SO(3)$ rotations in the triads. Recall
that a triad $e_a^i$ and a rotated triad $e^{\prime
i}_a(x)={U^i}_j(x)\,e^j_a(x)$ define the same metric $q_{ab}(x)$,
with ${U^i}_j(x) \in SO(3)$ a local rotation. In order to account
for the extra symmetry, there will be extra constraints (first
class) that will get rid of the extra degrees of freedom
introduced. Let us now introduce the densitized triad as follows:
\be
\tilde{E}^a_i=\half\,\epsilon_{ijk}\tilde{\eta}^{abc}\,e^j_b\,e^k_c
\ee
where $\tilde{\eta}^{abc}$ is the naturally defined levi-civita
density one antisymmetric object. Note that
$\tE^a_i\tE^b_j\delta^{ij}=q\,q^{ab}$.

Let us now consider the canonical variables. It turns out that the
canonical momenta to the densitized triad $\tE^a_i$ is closely
related to the extrinsic curvature of the metric, \be
K^i_a=\frac{1}{\sqrt{{{\rm
det}(\tE)}}}\;\delta^{ij}\tE^b_j\,K_{ab} \ee For details see
\cite{Perez:2004hj}. Once one has enlarged the phase space from
the pairs $(q_{ab},\tilde{\pi}^{ab})$ to $(\tE^a_i,K_b^j)$, the
next step is to perform the canonical transformation to go to the
$Ashtekar-Barbero$ variables. First we need to introduce the so
called {\it spin connection} $\G^i_a$, the one defined by the
derivative operator that annihilates the triad $e_a^i$ (in
complete analogy to the Christoffel symbol that defined the
covariant derivative $D_a$ killing the metric). It can be inverted
from the form,
\be
\partial_{[a}e^i_{b]} +{\epsilon^i}_{jk}\,\G^j_a\, e^k_b=0
\ee This can be seen as an extension of the covariant derivative
to objects with mixed indices. The key to the definition of the
new variables is to combine these two objects, namely the spin
connection $\G$ with the object $K^i_a$ (a tensorial object), to
produce a new connection
\be {}^{\gamma}\!A_a^i:=\G^i_a+\gamma\,K^i_a
\ee
This is the {\it Ashtekar-Barbero Connection}. Similarly, the
other conjugate variable will be the rescaled triad,
\be
{}^{\gamma}\!\tE^a_i=\tE^a_i/\gamma
\ee
Now, the pair $({}^{\gamma}\!A_a^i, {}^{\gamma}\!\tE^a_i)$ will
coordinatize the new phase space $\G_\gamma$. We have emphasized
the parameter $\g$ since this labels a one parameter family of
different classically equivalent theories, one for each value of
$\gamma$. The real and positive parameter $\g$ is known as the
Barbero-Immirzi parameter \cite{barbero,Immirzi}. In terms of
these new variables, the canonical Poisson brackets are given by,
\be \{ {}^{\gamma}\!A_a^i(x),
{}^{\gamma}\!\tE^b_j(y)\}=\kappa\,\delta^b_a\,\delta^i_j\,\delta^3(x,y)\,
. \ee and, \be \{ {}^{\gamma}\!A_a^i(x),
{}^{\gamma}\!A_b^j(y)\}=\{{}^{\gamma}\!\tE^a_i(x),
{}^{\gamma}\!\tE^b_j(y)\}=0 \ee Let us summarize. i) We started
with the geometrodynamical phase space where the configuration
space was taken to be the space of 3-Riemannian metrics, ii)
enlarged it by considering triads instead of metrics, and iii)
performed a canonical transformation, and changed the role of the
configuration variable; the connection ${}^{\gamma}\!A_a^i$ (that
has the information about the extrinsic curvature) is now the
configuration variable and the (densitized) triad is regarded as
the canonical momenta. The connection ${}^{\gamma}\!A_a^i$ is a
true connection since it was constructed by adding to the spin
connection $\G^i_a$ a tensor field, that yields a new connection.
We started by considering connection taking values in $so(3)$, the
Lie algebra of $SO(3)$, but since as Lie algebras it is equivalent
to $su(2)$, we will take as the gauge group the simply connected
group $SU(2)$ (which will allow to couple fermions when needed).
The phase space $\G_\g$ is nothing but the phase space of a
$SU(2)$ Yang-Mills theory.

Let us now write down the constraints in terms of the new phase
space variables. The new constraint that arises because of the
introduction of new degrees of freedom takes a very simple form,
\be G_i={\cal D}_a\,\tE^a_i\approx 0 \ee
that is, it has the structure of Gauss' law in Yang-Mills theory
and that is the name that has been adopted for it. We have denoted
by ${\cal D}$ the covariant defined by the connection
${}^{\gamma}\!A_a^i$, such that ${\cal
D}_a\tE^a_i=\partial_a\tE^a_i+{\epsilon_{ij}}^{k}\;{}^{\gamma}
\!A_a^j\tE^a_k$. The vector and scalar constraints now take the
form, \be V_a=F^i_{ab}\,\tE^b_i-(1+\g^2)K^i_a\,G_i\approx 0 \ee
where
$F^i_{ab}=\partial_{a}{}^{\gamma}\!A_b^i-\partial_{b}{}^{\gamma}\!A_a^i+
{\epsilon^i}_{jk}\;{}^{\gamma}\!A_a^j\,{}^{\gamma}\!A_b^k$ is the
curvature of the connection ${}^{\gamma}\!A_b^j$. The other
constraint is,
\be \S=\frac{\tE^a_i\tE^b_j}{\sqrt{{\rm
det}(\tE)}}\,\left[{\epsilon^{ij}}_k\,F^k_{ab}-2(1+\gamma^2)
\,K^i_{[a}K^j_{b]}\right] \approx 0 \ee
Note incidently that if $(1+\gamma^2)=0$, the constraints would
simplify considerably. That was the original choice of Ashtekar,
that rendered the connection complex (and thus the corresponding
gauge group non-compact), and it had a nice geometrical
interpretation in terms of self-dual fields. Historically the
emphasis from complex to real connections (and a compact group as
a consequence) was due to the fact that in this case the
mathematical well defined construction of the Hilbert space has
been completed, whereas the non-compact case remains open. For
more details see \cite{AL:review} and specially Sec 3.2 of
\cite{Perez:2004hj}.

The next step is to consider the right choice of variables, now
seen as functions of the phase space $\G_\g$ that are preferred
for the non-perturbative quantization we are seeking. As we shall
see, the guiding principle will be that the functions (defined by
an appropriate choice of smearing functions) will be those that
can be defined without the need of a background structure, i.e. a
metric on $\Sigma$.

\section{Holonomies and Fluxes}
\label{sec:3}

\noindent We have seen that the classical setting for the
formulation of the theory is the phase space of a Yang-Mill
theory, with extra constraints. Since the theory possesses these
constraints, the strategy to be followed is to quantize first and
then to impose the set of constraints as operators on a Hilbert
space. This Hilbert space is very important since it will be the
home where the imposition of the constraints will be implemented
and its structure will have some physical relevance. This is known
as the Kinematical Hilbert Space ${\cal H}_{\rm kin}$. One of the
main achievements of LQG is that this space has been rigourously
defined, something that was never done in the old geometrodynamics
program.

The main objective of this section is to motivate and
construct the classical algebra of observables that will be the
building blocks for the construction of the space ${\cal H}_{\rm
kin}$. In QFT one could say that there are two important choices
when quantizing a classical system. The first one is the choice of
the algebra of observables, consisting of two parts, the choice of
variables, and of functions thereof. The second choice is a
representation of this chosen classical algebra into a Hilbert
space. Both steps normally involve ambiguities. This is also true
of the phase space we are starting with. Remarkably for our case,
the physical requirements of background independence and
diffeomorphism invariance will yield a choice of variables and of
a representation that is in a sense, unique. The infinite freedom
one is accustomed to in ordinary QFT is here severely reduced.
Background independence and diffeomorphism invariance impose very
strong conditions on the possible quantum theories.

Let us start by considering the connection $A_a^i$ (from now on we
shall omit the $\gamma$ label). The most natural object one can
construct from a connection is a holonomy $h_\a(A)$ along a loop
$\a$. This is an element of the gauge group $G=SU(2)$ and is
denoted by,
\be h_\a(A)={\cal P}\,\exp\,\left( \oint_\a A_a\,\d s^a \right)
\ee
The path-order exponential of the connection. Note that for
notational simplicity we have omitted the `lie-algebra indices'.
The connection as an element of the lie algebra, in the
fundamental representation, should be written as
$A_a^i(\tau_{{}i})^A_B$, with $A,B=1,2$ the $2\times 2$-matrix
indices of the Pauli matrices $\tau_{{}i}$. From the holonomy, it
is immediate to construct a gauge invariant function by taking the
trace arriving then at the Wilson loop $T[\a] := \frac{1}{2}\,
{\rm Tr}\,\, {\cal P}\,\exp\,(\oint_\a A_a\, \d s^a)$.

Several remarks are in order. i) When seen as a smearing of the
connection, it is clear that the holonomy represents a one
dimensional smearing (as compared with, say, the three dimensional
smearings $\int A_a^i(x)g^a_i(x)\,\d^3 x$ used in ordinary QFT);
ii) The loop $\a$ can be seen as a label, but the holonomy {\it
is} a function of the connection $A^i_a$; iii) Even when the
holonomies are functions of the connection $A$, they only ``prove"
the connection along the loop $\a$. In order to have a useful set
of functions that can {\it separate points} of the space of smooth
connections ${\cal A}$, one needs to consider for instance Wilson
loop functions along {\it all} possible loops on $\Sigma$. The
algebra generate by such functions is called the {\it holonomy
algebra} ${\cal HA}$.

In order to implement the idea that one should look for
generalized notions that will replace the loops, let us consider
the most obvious extension. In recent years the emphasis has
shifted from loops to consider instead closed graphs $\Upsilon$,
that consist of $N$ edges  $e_I$ ($I=1,2,\ldots, N$), and $M$
vertices $v_\mu$, with the restriction that there are no edges
with `loose ends'. As a aside remark one should mention that every
graph $\Upsilon$ can be decomposed in independent loops $\a_i$
based at a chosen vertex $v$. Given  a graph $\Upsilon$, one can
consider the parallel transport along the edges $e_I$, the end
result is an element of the gauge group $g_I=h(e_I)\in G$ for each
such edge. One can then think of the connection $A^i_a$ as a map
from graphs to $N$-copies of the gauge group: $A^i_a:
\Upsilon\rightarrow G^N$. Furthermore, one can think of ${\cal
A}_{\Upsilon}$ as the configuration space  for the graph
$\Upsilon$, that is homeomorphic to $G^N$ (one can regard ${\cal
A}_{\Upsilon}$ as the configuration space of the `floating
lattice' gauge theory over $\Upsilon$).

Once we have recognized that one can associate a configuration
space for all graphs, one should not loose perspective that the
relevant classical configuration space is still the space ${\cal
A}$ of all (smooth) connections $A_a^i$. What we are doing at the
moment is to construct relevant configuration functions, making
use of the graphs and the space ${\cal A}_{\Upsilon}$. In
particular, what we need is to consider generalizations of the
Wilson loops $T[\a]$ defined previously. As we have mentioned
before, every graph $\Upsilon$ can be decomposed into independent
loops $\a_i$ and the corresponding Wilson loops $T[\a_i]$ are a
particular example of functions defined over ${\cal
A}_{\Upsilon}$. What we shall consider as a generalization of the
Wilson loop are {\it all} possible functions defined over ${\cal
A}_{\Upsilon}$ (in a sense the Wilson loops generate these
functions, but are overcomplete). Thus, a function
$c:G^N\rightarrow \mathbb{C}$ defines a {\it cylindrical} function
$C_\Upsilon$ of the connection $A$ as,
  \be
C_\Upsilon:= c(h(e_1),h(e_2),\ldots,h(e_N))
  \ee
By considering all possible functions $c$ and all possible
embedded graphs $\Upsilon$, we generate the algebra of functions
known as Cyl (it is closely related to the holonomy algebra, and
it can be converted into a $C^*$-algebra $\overline{\rm Cyl}$, by
suitable completion).

Even when we shall consider in what follows the full algebra Cyl,
one should keep in mind that the basic objects that build it are
precisely the holonomies, functions of the connection smeared
along one dimensional objects. Let us now discuss why this choice
of configuration functions is compatible with the basic guiding
principles for the quantization we are building up, namely
diffeomorphism invariance and background independence. Background
independence is clear since there is no need for a background
metric to define the holonomies. Diffeomorphism invariance is a
bit more subtle. Clearly, when one applies a diffeomorphism
$\phi:\Sigma\rightarrow \Sigma$, the holonomies transform in a
covariant way
\be
  \phi_* \cdot h(e_I)=h(\phi^{-1}\cdot e_I)\, ,
\ee
that is, the diffeomorphism acts by moving the edge (or loop). How
can we then end up with a diffeo-invariant quantum theory? The
strategy in LQG is to look for a {\it diffeomorphism invariant}
representation of the diffeo-covariant configuration functions. As
we shall see later, this has indeed been possible and in a sense
represents the present `success' of the approach.

Let us now consider the functions depending of the momenta that
will be fundamental in the (loop) quantization. The basic idea is
again to look for functions that are defined in a background
independent way, that are natural from the view point of the
geometric character of the object (1-form, 2-form, etc), and that
transform covariantly with respect to the gauge invariances of the
theory. Just as the the connection $A_a^i$ can be identified with
a one form that could be integrated along a one-dimensional
object, one would like analyze the geometric character of the
densitised triad $\tE$ in order to naturally define a smeared
object. Recall that the momentum is a density-one vector field on
$\Sigma$, $\tE^a_i$ with values in the dual of the lie-algebra
$su(2)$. In terms of its tensorial character, it is naturally dual
to a (lie-algebra valued) two form,
\be
 E_{ab\;i}:=\half\, \teta_{abc}\,\tE^c_i
\ee
where $\teta_{abc}$ is the naturally defined Levi-Civita symbol.
It is now obvious that the momenta is crying to be integrated over
a two-surface $S$. It is now easy to define the objects
\be
 E[S,f]:=\int_S \;E_{ab\;i}\,f^i\,\d S^{ab}\, ,
\ee
where $f^i$ is a lie-algebra valued smearing function on $S$. This
`Electric flux' variable does not need a background metric to be
defined, and it transforms again covariantly as was the case of
the holonomies. The algebra generated by holonomies and flux
variables is known as the {\it Holonomy-Flux} algebra ${\cal HF}$.

Perhaps the main reason why this Holonomy-Flux  algebra
${\cal HF}$ is interesting, is the way in which the basic
generators interact, when considering the classical (Poisson)
lie-bracket. First, given that the configuration functions depend
only on the connection and the connections Poisson-commute, one
expects that $\{T[\a],T[\beta]\}=0$ for any loops $\a$ and
$\beta$. The most interesting poisson bracket one is interested in
is the one between a configuration and a momenta variable,
\be \{T[\a],E[S,f]\}= \kappa\,\sum_\mu
f^i(v)\,\iota(\a,S|_{v})\,{\rm Tr} \,(\tau_i\, h(\a))
\label{poissonB}
 \ee
where the sum is over the vertices $v$ and
$\iota(\a,S|_{v})=\pm 1$ is something like the
intersection number between the loops $\a$ and the surface $S$ at
point $v$. The sum is over all intersection of the loop $\a$
and the surface $S$. The most important property of the Poisson
Bracket is that it is completely topological. This has to be so if
we want to have a fully background independent classical algebra
for the quantization.

A remark is in order. The value of the constant $\iota|_{v}$
depends not only on the relative orientation of the tangent vector
of the loop $\a$ with respect to the orientation of $\Sigma$ and
$S$, but also on a further decomposition of the loop into edges,
and whether they are `incoming' or `outgoing' to the vertex
$v$. The end result is that is we have, for simple
intersections, that the number $\iota|_{v}$ becomes insensitive
to the `orientation'. This is different to the $U(1)$ case where
the final result {\it is} the intersection number. For details see
\cite{ACZ} (and note the difference with the claims in
\cite{Nicolai:2005mc}).

Let us now consider the slightly more involved case of a
cylindrical function $C_\Upsilon$ that is defined over a graph
$\Upsilon$ with edges $e_I$, intersecting the surface $S$ at
points $p$. We have then,
\be \{ C_\g, \, E[S,f] \} = \frac{\kappa}{2} \sum_{p}\,
\sum_{I_p}\, \iota({I_p})\, f^i(p)\, X^i_{I_p}\cdot c
\label{poissonbb}
 \ee
where the sum is over the vertices $p$ of the graph that lie on
the surface $S$, $I_p$ are the edges starting or finishing in $p$
and where $X^i_{I_P}\cdot c$ is the result of the action of the
$i$-th left (resp. right) invariant vector field on the $I_p$-th
copy of the group if the $I_p$-th edge is pointing away from
(resp. towards) the surface $S$.  Note the structure of the right
hand side. The result is non-zero only if the graph $\Upsilon$ used in
the definition of the configuration variable $C_\Upsilon$ intersects the
surface $S$ used to smear the triad. If the two intersect, the
contributions arise from the action of right/left invariant vector
fields on the arguments of $c$ associated with the edges at the
intersection.

Finally, the next bracket we should consider is between two
momentum functions, namely $\{E[S,f],E[S^\prime,g]\}$. Just as in
the case of holonomies, these functions depend only on one of the
canonical variables, namely the triad $\tE$. One should then
expect that their Poisson bracket vanishes. Surprisingly, this is
{\it not} the case and one has to appropriately define the correct
algebraic structure\footnote{The end result is that one should not
regard $E[S,f]$ as phase space functions subject to the ordinary
Poisson bracket relations, but rather should be viewed as arising
from vector fields $X^\alpha$ on ${\cal A}$. The non-trivial
bracket is then due to the non-commutative nature of the
corresponding vector fields. This was shown in \cite{ACZ} where
details can be found}.

We have arrived then to the basic variables that will be used in the
quantization in order to arrive at LQG. They are given by,
\be
h(e_I)\qquad {\rm Configuration \; function}
\ee
and
\be
  E[S,f]\qquad {\rm Momentum \; function},
\ee subject to the basic Poisson bracket relations given by
Eqs.~(\ref{poissonB}) and (\ref{poissonbb}). In the next section
we shall take the Holonomy-Flux algebra ${\cal HF}$ as the
starting point for the quantization.

\section{The Hilbert Space}
\label{sec:4}

\noindent In this section we shall first outline the quantization
strategy to arrive at the kinematical Hilbert space $\H_{\rm
kin}$. Later on, we shall give certain details of the several
parts that arise in the construction. As we have emphasized, the
Kinematical Hilbert space is the starting point for the program of
implementing the constraints as quantum operators. We shall
consider briefly that issue later on.

\subsection{General Considerations}

The strategy is to build the Hilbert space as in the ordinary
Schr\"odinger representation, where states are to be represented
by wave-functions of the configuration space. Recall that in the
present approach, one has decided that the space ${\cal A}$ is the
space of (classical) configurations. The natural strategy is then
to consider wave functions $\Psi$ of the form, \be \Psi=\Psi(A)
\ee Following the analogy, one should expect that the Hilbert
space will consist of wave-functions that are square integrable.
This means that one has to introduce a measure $\d\mu$ on the
space of connections, in order to define the inner product
heuristically as,
\be \langle\Phi|\Psi\rangle=\int\d\mu\;\overline{\Phi}\,\Psi\, \ee
The Hilbert space would be denoted then as $\H_{\rm kin}=L^2({\cal
A},\d\mu)$. In this configuration representation one expects that
the connection and any function of it will act as a multiplication
operator, and that the momenta will be represented as a
derivation, with possible a correction term\footnote{For a
discussion of the corresponding Schr\"odinger representation for a
scalar field see \cite{CCQ}.}.

There are several different technical issues that need to be
properly addressed in order to complete the construction as we
have outlined it. Basically what needs to be done is to give
mathematical meaning to the different aspects of the kinematical
Hilbert space. Let us then make some general remarks for each of
them.

i) {\it Configuration space}. From the experience that has been
gained from the scalar field case, we know that the space where
the wave functions have support is a much larger (functional)
space as the space of classical configurations. In the scalar
case, the classical configuration space involves smooth fields,
whereas the {\it quantum configuration space} is made of tempered
distributions (dual to the Schwarz space). In our case we expect
that the quantum configuration space $\overline{\cal A}$ will also
be an extension of the classical space. This is indeed the case as
was shown in the mid-nineties by Ashtekar and collaborators. There
are several characterizations of this space (that can be found in
\cite{AL:review, Thiemann:2001yy, Velhinho}), but here we shall
mention only two of them (for a third one see below). The first
one is to recall that one could think of a connection as an
operator that acts on 1-dimensional objects (edges) and gives a
group element. Certainly, a smooth connection has this property.
It turns out that a {\it generalized connection}, an element of
the quantum configuration space $\overline{\cal A}$, is precisely
{\it any} such map (satisfying some tame conditions such as the
composition: $h(e_1\circ e_2)=h(e_1)\cdot h(e_2)\,;,\;\forall\;
e_I$). In particular, what is dropped is any notion of continuity
of the connection. The other characterization is more algebraic.
It is based on the observation that the (Abelian) algebra of
cylindrical functions Cyl can be extended to a $C^*$-algebra
$\overline{\rm Cyl}$ with unit. Standard theorems on
representations of such algebras tell us that those algebras can
be viewed as the space of continuous functions over a compact
space $\Delta$, called the spectrum of the algebra. It turns out
that $\overline{\cal A}$ is precisely the spectrum of the algebra
$\overline{\rm Cyl}$.

ii) {\it Measure.} A extremely important issue in the choice of
quantum theory is the inner product on the Hilbert space. In the
case of wave-functions this implies defining a measure on the
configuration space $\overline{\cal A}$. At his point one should
require that the measure --responsible in a sense for the
resulting quantum representation-- be diffeomorphism invariant.
This would implement naturally the intuitive notion that the
theory should be spatially diffeo-invariant. The construction of a
measure with this property is one of the main results of the
program since it was generally thought that   no such measures
existed for gauge theories. The diffeo-invariant measure is known
as the Ashtekar-Lewandowski measure $\mu_{\rm AL}$.

 iii) {\it Momentum
operators.} The other important part of the quantum representation
is of course, the way in which the functions $E[S,f]$ are promoted
to operators on the Hilbert space $\H_{\rm kin}$. Intuitively,
whenever the wave function $\Psi$ is a function of the
configuration variable (the connection), the conjugate variable
acts as derivation $\hat{E^a}=\delta/\delta A_a$. However, one
should also consider the possibility of adding a multiplicative
term to the derivative. This term would commute with the
(multiplicative) configuration operator $\hat{h}(e_I)$, so the
commutation relations would be satisfied. This extra term  would
also account, for instance, for the existence of a non-trivial
measure (See \cite{CCQ} for a discussion in the scalar case).
Therefore, the choice of measure and the representation of the
momentum operator are intertwined, and one should ensure that the
measure not only be diffeomorphism invariant, but should also
``support" the triad. A systematic study of these issues initiated
by Sahlmann has yielded a uniqueness result: The only
diffeo-covariant representation of the Holonomy-Flux algebra
${\cal HF}$ is given by the Ashtekar-Lewandowski representation
\cite{LOST}.

iv) {\it Hybrid variables.} Recall that the basic functions we
have chosen for the quantization are the pairs $(h(e_I),E[S,f])$.
These generators of the classical algebra to be quantized (the
Holonomy-Flux algebra), are not canonical. One of them, namely the
holonomy, is a exponentiated version of the connection, whereas
the electric flux is linear in the triad. This would be equivalent
to a choice $U(\lambda)=\exp[i\lambda q]$ and $\pi(\nu)=\nu\,p$,
for a quantum mechanical system with phase space $\G=(q,p)$. This
means that the basic functions of phase space are neither of the
canonical type $(q,p)$, nor of the Weyl form (both exponentiated);
the quantization based on this functions is in a strict sense
non-canonical. In finite dimensions the Stone von Neumann theorem
assures us that the Weyl relations are equivalent to the Canonical
Commutation Relations, provided the quantum representations are
{\it regular}\footnote{For a nice discussion see for instance
\cite{wald2}.}. For a representation to be regular means that the
exponentiated form of the operators --which are unitary-- are
continuous with respect to the parameters ($\lambda$ in the
$U(\lambda)$ above). One of the main features of the loop
quantization is that the exponentiated variable, the holonomy, has
a well defined associated operator $\hat{h}(e_I)$ on $\H_{\rm
kin}$ that is however, discontinuous. Thus, the operator
$\hat{A}_i^a$ does {\it not} exist! In the finite-dimensional
system example, the equivalent statement would be to say that
$\hat{U}(\lambda)$ is well defined but $\hat{q}$ is not. Such
non-regular representations in quantum mechanics {\it do} exist
\cite{poly} and form the basic starting point for Loop Quantum
Cosmology \cite{LQC,bojo}.

\subsection{Ashtekar-Lewandowski Hilbert Space}

Let us now consider the particular representation that defines
LQG. As we have discussed before, the basic observables are
represented as operators acting on wave functions
$\Psi_\Upsilon(\overline{A})\in{\cal H}_{\rm kin}$ as follows:
\be \hat{h}(e_I)\cdot
\Psi(\overline{A})=\left(h(e_I)\;\Psi\right)(\bar{A})
 \ee
and
\be\label{11} \hat{E}[S,f]\cdot\Psi_\Upsilon(\overline{A})=i\hbar\,
\{ \Psi_\Upsilon, \, {}^2\!E[S,f] \} = i\frac{\ell^2_{\rm P}}{2}
\sum_{p}\, \sum_{I_p}\, \kappa({I_p})\, f^i(p)\, X^i_{I_p}\cdot
\psi
\ee
where $\ell^2_{\rm P}=\kappa\,\hbar=8\pi\,G\,\hbar$,
the Planck area is giving us
the scale of the theory (recall that the Immirzi parameter $\g$
does not appear in the basic Poisson bracket, and should therefore
not play any role in the quantum representation). Here we have
assumed that a cylindrical function $\Psi_\Upsilon$ on a graph
$\Upsilon$ is an element of the Kinematical Hilbert space (which
we haven't defined yet!). This implies one of the most important
assumption in the loop quantization prescription, namely, that
objects such as holonomies and Wilson loops that are smeared in
one dimension are well defined operators on the quantum
theory\footnote{this has to be contrasted with the ordinary Fock
representation where such objects do not give raise to well
defined operators on Fock space. This implies that the loop
quantum theory is qualitatively different from the standard
quantization of gauge fields.}.

 Let us now introduce the third
characterization of the quantum configuration space which will be
useful for constructing the Hilbert space ${\cal H}_{\rm kin}$.
The basic idea for the construction of both the Hilbert space
${\cal H}_{\rm kin}$ (with its measure) and the quantum
configuration space $\bar{\cal A}$, is to consider the projective
family of {\it all} possible graphs on $\Sigma$. For any given
graph $\Upsilon$, we have a configuration space ${\cal
A}_\Upsilon=(SU(2))^N$, which is $n$-copies of the (compact) gauge
group $SU(2)$. Now, it turns out that there is a preferred
(normalized) measure on any compact semi-simple Lie group that is
left and right invariant. It is known as the Haar measure
$\mu_{\rm H}$ on the group. We can thus endow ${\cal A}_\Upsilon$
with a measure $\mu_\Upsilon$ that is defined by using the Haar
measure on all copies of the group. Given this measure on ${\cal
A}_\Upsilon$, we can consider square integrable functions thereon
and with them the graph-$\Upsilon$ Hilbert space ${\cal
H}_\Upsilon$, which is of the form:
\be
 {\cal H}_\Upsilon=L^2({\cal A}_\Upsilon,\d\mu_\Upsilon)
\ee
If we were working with a unique, fixed graph $\Upsilon_0$ (which
we are not), we would be in the case of a lattice gauge theory on
an irregular lattice. The main difference between that situation
and LQG is that, in the latter case, one is considering all graphs
on $\Sigma$, and one has a family of configurations spaces $\{
{\cal A}_\Upsilon \,/ \Upsilon \,{\rm a\,graph\,in\,} \Sigma\}$,
and a family of Hilbert spaces $\{ {\cal H}_\Upsilon \,/ \Upsilon
\,{\rm a\,graph\,in\,} \Sigma\}$. In order to have  consistent
families of configuration  and Hilbert spaces one needs some
conditions. In particular, there is a notion of when a graph
$\Upsilon$ is `larger' than $\Upsilon'$. We say that if $\Upsilon$
contains $\Upsilon'$ then $\Upsilon > \Upsilon'$.\footnote{by
containing we mean that the larger graph can be obtained from the
smaller by adding some edges or by artificially dividing the
existing edges (by proclaiming that an interior point $p\in e_I$
is now a vertex of the graph).}

Given this (partial) relation ``$>$", we have a corresponding
projection $P_{\Upsilon,\Upsilon'}:\Upsilon\mapsto\Upsilon'$,
which in turn induces a projection operator for configuration
spaces $P:{\cal A}_\Upsilon\mapsto {\cal A}_\Upsilon'$ and an
inclusion operator for Hilbert spaces $\iota:{\cal
H}_\Upsilon'\mapsto {\cal H}_\Upsilon$. The consistency conditions
that need to be imposed are rather simple. The intuitive idea is
that, if one has a graph $\Upsilon_o$ and a function thereon
$\psi_\Upsilon[A]$ one should be able to consider larger graphs
$\Upsilon_i$ where the function $\psi$ be well defined; one should
be able to talk of the `same function', even when defined on a
larger graph, and more importantly, its integral (and inner
product with other functions) should be independent of the graph
$\Upsilon_i$ we have decided to work on.

We are now in the position of giving a heuristic definition of the
configuration space $\overline{\cal A}$ and ${\cal H}_{\rm kin}$:
The quantum configuration space $\overline{\cal A}$ is the
configuration space for the ``largest graph"; and similarly, the
kinematical Hilbert space ${\cal H}_{\rm kin}$ is the largest
space containing all Hilbert spaces in $\{ {\cal H}_\Upsilon \,/
\Upsilon \,{\rm a\,graph\,in\,} \Sigma\}$. Of course, this can be
made precise mathematically, where the relevant limits are called
{\it projective}. For details see \cite{alm2t,AL:review} and
\cite{Thiemann:2001yy}. The Ashtekar-Lewandowski  measure
$\mu_{\rm AL}$ on ${\cal H}_{\rm kin}$ is then the measure whose
projection to any ${\cal A}_\Upsilon$ yields the corresponding
Haar measure $\mu_\Upsilon$. The resulting Hilbert space can thus
be written as
$$
{\cal H}_{\rm kin}=L^2(\overline{\cal A},\d\mu_{\rm AL})
$$
Let us now see how it is that the cylindrical functions
$\Psi_\Upsilon\in$ Cyl belong to the Hilbert space of the theory.
Let us consider a cylindrical function $\Psi_\Upsilon$ defined on
the space ${\cal A}_\Upsilon$. If it is continuous, then it is
bounded (since ${\cal A}_\Upsilon$ is compact), and thus it is
square integrable with respect to the measure $\mu_\Upsilon$.
Therefore, $\Psi_\Upsilon \in {\cal H}_\Upsilon$. Finally, we have
the inclusion between Hilbert spaces given by ${\cal H}_\Upsilon
\subset {\cal H}_{\rm kin}$ which implies that the cylinder
function $\Psi_\Upsilon$ indeed belongs to the kinematical Hilbert
space.

Let us now see how this inner product works for known functions
such as Wilson loops. Consider $T[\a]$ and $T[\beta]$ two Wilson
loops with $\a \neq \beta$ (and nonintersecting, for simplicity).
Each loop can be regarded as a graph on itself, with only an edge
and no vertex (or many bi-valent vertices by artificially
declaring them to be there). In order to take the inner product
between these two function which looks like
\be \langle\, T[\a]\,|\,T[\beta]\, \rangle_{\rm
kin}=\int_{\overline{\cal A}} \d\mu_{\rm AL}\;
\overline{T[\a]}\;T[\beta]\label{inn-pro}
 \ee
we have to construct a graph $\Upsilon$ that contains both $\a$
and $\beta$. This is rather easy to do and in fact there is large
freedom in doing that. The end result should be independent of the
particular choice. The simplest possibility is to take an edge $e$
that connects {\it any} point of $\alpha$ with any point of
$\beta$ (and for simplicity it does not intersect neither loop at
another point). The resulting graph has now three edges
$(\a,\beta,e)$, so we can construct its configuration space to be
${\cal A}_\Upsilon=(SU(2))^3=$. The Wilson loops are now very
simple cylinder functions on ${\cal A}_\Upsilon$ (or rather, the
induced function by the inclusion of $\a$ into $\Upsilon$).
$T[\a]$ has associated a function $t_\a:(SU(2))^3\rightarrow
\mathbb{C}$ that only depends on the first argument (corresponding
to $\a$), whereas $T[\beta]$ has a function $t_\beta$ that depends
only on the second argument .Its functional dependence (the same
for $T[\a]$ and $T[\beta]$), as a function of the holonomy in each
edge, is very simple:
$$
T[\a]=t_\a(h(\a),h(\beta),h(e))=\half{\rm Tr}(h(\a))\, .
$$
Given that we are integrating a function
($\overline{T[\a]}\;T[\beta]$) that is defined on the graph
$\Upsilon=\a\cup\beta\cup e$, and that the functional integral
over the full space $\overline{\cal A}$ reduces to an integral
over the minimal graph were the function is defined, the inner
product (\ref{inn-pro}) can then be rewritten as,
\be \langle\, T[\a]\,|\,T[\beta]\, \rangle_{\rm kin}=\int_{G_1}
\d\mu_{\rm H}\; \overline{T[\a]}\;\int_{G_2} \d\mu_{\rm
H}\;T[\beta]\; \int_{G_3} \d\mu_{\rm H} \label{inn-pro2}
 \ee
Each integral is performed over each copy of the gauge group,
where $G_1$ denotes the first entry in the functions $t_i$, namely
the holonomy along $\a$, and $G_2$ the holonomy along $\beta$. The
third copy of the group, that associated to $e$ has the property
that the function we are integrating has no dependence on it, thus
one is only integrating the unit function. We are assuming that
$\mu_H$ is normalized so this integral is equal to one. The
question now reduces to that of calculating the integral:
\be \int_{G_1} \d\mu_{\rm
H}\;\overline{T[\a]}=\frac{1}{2}\int_{G_1} \d\mu_{\rm H}\;
\overline{{\rm Tr}(h(\a))}\label{vev} \ee It turns out that the
Haar measure is such that this integral vanishes. Thus, the inner
product (\ref{inn-pro}) is zero, for any two (different) loops
$\alpha$ and $\beta$ and for any edge $e$ connecting them. Note
that if we had assumed from the very beginning that the loops $\a$
and $\beta$ are equal, then there would be no need to define a
connecting edge and the inner product would be given by only one
integral $ \int_{G_1} \d\mu_{\rm
H}\;\overline{T[\a]}\,T[\a]=\frac{1}{2}\int_{G_1} \d\mu_{\rm H}\;
|{\rm Tr}(h(\a))|^2\neq 0$

Several remarks are in order:

i) The fact that the inner product between any two Wilson loop
functions vanishes is a clear signature that the measure $\mu_{\rm
AL}$ is diffeomorphism invariant. More precisely, a diffeomorphism
$\phi$ induces a transformation $U(\phi):{\cal H}\rightarrow{\cal
H}$ given by:
$U(\phi)\cdot\Psi_\Upsilon=\Psi_{(\phi^{-1}\cdot\Upsilon)}$. The
measure $\mu_{\rm AL}$ is such that the operator $U(\lambda)$ is
unitary. Thus, diffeomorphism are unitarily implemented in this
quantum theory.

ii) The integral (\ref{vev}) can be interpreted as the vacuum
expectation value of the operator $\hat{T}[\a]$; The ``vacuum"
$\Psi_0$ in this representation is simply the unit function, so we
have:
$$
\langle\,\hat{T}[\a]\,\rangle_0=
\int_G T[\a]\,\d\mu_{\rm H}=0\, .
$$
We can then conclude that in the LQG representation, the vacuum
expectation value of all Wilson loops vanishes exactly.

iii) As we mentioned before, the kinematical Hilbert space ${\cal
H}_{\rm kin}$ can be regarded as the ``largest" Hilbert space by
combining the Hilbert spaces over all possible graphs. We shall
use the following symbol to denote that idea. We write then:
$$
{\cal H}_{\rm kin}=\otimes_{\Upsilon}\,{\cal H}_\Upsilon
$$
Note that since the space of graphs $\Upsilon$ is uncountable, the
Hilbert space ${\cal H}_{\rm kin}$ is non-separable. In the next
part we shall se how a convenient choice of basis for each Hilbert
space ${\cal H}_\Upsilon$ will allow us to have a full
decomposition of the Hilbert space.

Let us then recall what is the structure of simple states in the
theory. The vacuum or `ground state' $|0\rangle$ is given by the
unit function. One can then create excitations by acting via
multiplication with holonomies or Wilson loops. The resulting
state $|\a\rangle=\hat{T}[\a]\cdot|0\rangle$ is an excitation of
the geometry but only along the one dimensional loop $\a$. Since
the excitations are one dimensional, the geometry is sometimes
said to be {\it polymer like}. In order to obtain a geometry that
resembles a three dimensional continuum one needs aa huge number
of edges ($10^68$) and vertices.

\subsection{A choice of basis: Spin Networks}

The purpose of this part is to provide a useful decomposition of
the Hilbert space ${\cal H}_\Upsilon$, for all graphs. In practice
one just takes one particular graph $\Upsilon_o$ and works in that
graph. This would be like restricting oneself to a fixed lattice
and one would be working on the Hilbert space of a Lattice Gauge
Theory. Thus, all the results of this subsection are also relevant
for a LGT, but one should keep in mind that one is always
restricting the attention to a little part of the total Hilbert
space ${\cal H}_{\rm kin}$.

Let us begin by sketching the basic idea of what we want to do.
For simplicity, let us consider just one edge, say $e_i$. What we
need to do is to be able to decompose any function $F$ on $G$ (in
this case we only have one copy of the group), in a suitable
basis. In the simplest case, when the group is just $G=U(1)$ we
just have a circle. In this case, the canonical decomposition for
any function $F$ is given by the Fourier series: $F=\sum_n
A_n\,\e^{\ii \,n\,\theta}$, with $A_n=\frac{1}{2\pi}\int
F(\theta)\e^{-\ii n\theta}\d\theta$. As is well known, the
functions $\e^{\ii \,n\,\theta}$ represent a basis for any
function on the group $G=U(1)$, that is also orthonormal with
respect to the measure $\d\mu_{\rm H}=\frac{1}{2\pi}\d\theta$,
which is nothing but the Haar measure on this group. For
notational simplicity one can denote by $f_n(\theta)=\e^{\ii
\,n\,\theta}$, and they represent irreducible representations of
the group $U(1)$, labelled by $n$. Then the series looks like
$F(\theta)=\sum_n\,A_n\,f_n(\theta)$. Here we should think of
assigning, to each of the basis functions $f_n$ the trivial label
$n$ (for all integers), and the general function is a linear
combination of the $n$-labelled basis.

In the case of the group $G=SU(2)$, there is en equivalent decomposition of
a function $f(g)$ of the group ($g\in G$). It reads,
\be
 f(g)=\sum_j\sqrt{j(j+1)}\,f^{m m'}_j\stackrel{j}{\Pi}_{m m'}\;(g)
\ee where, \be f^{m m'}_j=\sqrt{j(j+1)}\int_G\d\mu_H
\stackrel{j}{\Pi}_{m m'}(g^{-1})\,f(g) \ee is the equivalent of
the Fourier component. When doing harmonic analysis on groups the
generalization of the Fourier decomposition is known as the
Peter-Weyl decomposition. The functions $\stackrel{j}{\Pi}_{m
m'}\,(g)$ play the role of the Fourier basis. In this case these
are unitary representation of the group, and the label $j$ labels
the irreducible representations. In the $SU(2)$ case with the
interpretation of spin, these represent the spin-$j$
representations of the group. In our case, we will continue to use
that terminology (spin) even when the interpretation is somewhat
different.

Given a cylindrical function
$\Psi_\Upsilon[A]=\psi(h(e_1),h(e_2),\ldots,h(e_N))$, we can then
write an expansion for it as,
\ba
\Psi_\Upsilon[A] &=& \psi(h(e_1),h(e_2),\ldots,h(e_N))\nonumber\\
& = & \sum_{j_1\cdots j_N}f^{m_1\cdots m_N,n_1\cdots
n_N}_{j_1\cdots j_N}
\;\phi^{j_1}_{m_1n_1}(h(e_1))\cdots\phi^{j_N}_{m_Nn_N}(h(e_N)),
\ea
where $\phi^{j}_{mn}(g)=\sqrt{j(j+1)}\,\stackrel{j}{\Pi}_{m
n}(g)$ is the normalized function satisfying
$$
\int_G \d\mu_{\rm H}\,\overline{\phi^{j}_{mn}(g)}\;\phi^{j'}_{m' n'}(g)=
\delta_{j,j'}
\delta_{m,m'}\delta_{n,n'}\, .
$$
The expansion coefficients can be obtained by projecting the state
$|\Psi_\Upsilon\rangle$,
\be
f^{m_1\cdots m_N,n_1\cdots n_N}_{j_1\cdots j_N} =\langle
\,\phi^{j_1}_{m_1n_1}\cdots\phi^{j_N}_{m_Nn_N}\;|\;\Psi_\Upsilon
\rangle
\ee
This implies that the products of components of irreducible
representations $\prod^{N}_{i=1}\phi^{j_i}_{m_i n_i}[h(e_i)]$
associated with the $N$ edges $e_I\in\Upsilon$, for all values of
spins $j$ and for $-j\leq m,n\leq j$ and for any graph $\Upsilon$,
is a complete orthonormal basis for ${\cal H}_{\rm kin}$. We can
the write, \be {\cal H}_\Upsilon=\otimes_j\;{\cal H}_{\Upsilon,j}
\ee where the Hilbert space ${\cal H}_{\a,j}$ for a single loop
$\a$, and a label $j$ is the familiar $(2j+1)$ dimensional Hilbert
space of a particle of `spin $j$'. For a complete treatment see
\cite{Perez:2004hj}.

We have been able to decompose the Hilbert space of a given graph,
using the Peter-Weyl decomposition theorem, but how can we make
contact with the so called spin networks? Let us for a moment
focus our attention to the $U(1)$ case. Then, for each edge one
could decompose any function in a Fourier series where the basis
was labelled by an integer, that correspond to all possible
irreducible unitary representations of the group. As we saw later,
one could think of a basis of the graph Hilbert by considering the
product over functions labelled by these representations. What one
can do it to associate a label $n_I$ for each edge, to denote the
function made out of those basis functions corresponding to the
chosen labels. A graph with the extra labelling, known as
dressing, can be given different names. In the $U(1)$ theory the
labelling denotes the possible electric flux, so the graph
$\Upsilon_{(n_1,n_2,\cdots,n_N)}$ with the labellings for each
edge is called a {\it flux network}. In the case of geometry with
group $SU(2)$, the graphs with labelling $j_I={\bf j}$ are known
as spin networks. As the reader might have noticed, in the
geometry case there are more labels than the spins for the edges.
Normally these are associated to vertices and are known as
intertwiners. This means that the Hilbert spaces ${\cal
H}_{\Upsilon,{\bf j}}$ is finite dimensional. Its dimension being
a measure of the extra freedom contained in the intertwiners. One
could then introduce further labelling $\bf l$ for the graph, so
we can decompose the Hilbert space as
\be
{\cal H}_\Upsilon=\otimes_j\;{\cal H}_{\Upsilon,{\bf
j}}=\otimes_{{\bf j},{\bf l}}\;{\cal H}_{\Upsilon,{\bf j},{\bf l}}
\ee
where now the spaces ${\cal H}_{\Upsilon,{\bf j},{\bf l}}$ are
one-dimensional. For more details see \cite{baez2}, Sec.~4.1.5 of
\cite{Perez:2004hj} and Sec.~4.2 of \cite{AL:review}.

Let us see the simplest possible example, the Wilson loop $T[\a]$.
In this case there is only one edge, and the function only
involves the lowest representation $j=1/2$. Furthermore, the
example is very easy since it is only the trace of the object
$\phi^{1/2}_{m n}$. In terms of labelling it corresponds to a
$(j=1/2)$ label on the loop. Since there are no vertices, there is
no choice for `intertwiner'. The next simplest example of a spin
network is for the simplest graph, namely a loops $\a$ but for
higher spin $j$ labellings. These functions correspond to taking
the trace of the holonomy around $\a$ but in a higher
representation $j$ of the group. In general a spin-$j$ Wilson loop
$T^j[\a]$ can be written in terms of the fundamental one $T[\a]$
as a polynomial expansion of degree $j$:
$T^j[\a]=\sum_{k=1,\ldots,j}A_k\,(T[\a])^k$, for some coefficients
$A_k$.

Two final remarks are in order. While it is true that spin
networks are naturally defined in terms of the harmonic analysis
on $G$, and due to its properties as an eigen-basis of important
operators (the subject of next chapter) have a special place in
the theory, they are not, by themselves {\it the} theory. Dirac's
transformation theory assures us that there are equivalent
descriptions for the quantum theory that are not given by these
functions. Another related issue is the existence of the so-called
{\it loop representation} of the theory. In its early stages,
non-perturbative quantum gravity (today known as LQG) was though
to have two equivalent representations, the connection
representation and the loop representation (precisely in the sense
of Dirac's transformation theory). When the structure of the
theory was properly understood and the Hilbert space was
characterized, it was clear that the loop representation as
originally envisioned was not well defined. The basic idea was to
have loops $\a$ as the argument for wave-functions. Thus, if one
had (in Dirac notation) a state $|\Psi\rangle$, the wave function
of the connection would be $\Psi[A]=\langle A|\Psi\rangle$. If one
had a `basis of loops' $\langle\a|$, one could imagine having
$\langle\a|\Psi\rangle$, the state in the loop representation.
Furthermore, there was a proposal for defining this state via the
`loop transform':
\be \psi[\a]=\langle\a|\Psi\rangle=\int {\cal D}A\;
\langle\a|A\rangle\,\langle A|\Psi\rangle := \int {\cal D}A\;
\overline{T}[\a,A]\,\Psi[A]
\ee
It was then implied that the Wilson loops were the kernel of the
transformation and were in a sense complete. From the discussion
of this section, it is clear that those expectations were
superseded. Namely, what we have seen is that loops by themselves
are not enough; we would loose information by only considering
loops\footnote{More precisely, when seen as graphs, loops are not
enough. What was previously done was to consider integer powers of
loops $\a^n=\a\circ\a\cdots\circ\a$ in the `basis' of loops.
Graphs (that only care about the image of the embedding, and where
$\a^2=\a$) together with spin and intertwiners represent a more
efficient (no redundancies) way of characterizing the {\it
discrete} basis of the theory.}.
 We need the extra information contained in the spin
networks, namely we need to consider all possible spins and
intertwiners. Instead, what one has to consider is the `spin network
transform',
\be \psi[\Upsilon_{{\bf j},{\bf l}}]=\langle \Upsilon_{{\bf
j},{\bf l}}|\Psi\rangle=\int \d\mu_{\rm AL} \;
\langle \Upsilon_{{\bf j},{\bf l}}  |A\rangle\,\langle A|\Psi\rangle
:= \int_{\bar{\cal A}}
\d\mu_{\rm AL}\; \overline{\cal N}[\Upsilon_{{\bf j},{\bf l}}
,A]\,\Psi[A]
\ee
Let us end this part with a two remarks regarding spin networks.
So far we have only considered cylindrical functions that satisfy
no further requirements. We know, on the other hand that in order
to have physical states we need to impose the quantum constraints
on the states. In particular Gauss' law implies that the states be
gauge invariant, namely, invariant under the action of (finite)
gauge transformations of the connection $A\mapsto
g^{-1}Ag+g^{-1}{\rm d}g$. This condition, when translated to the
language of holonomies, graphs and so on imposes some restrictions
on the cylinder functions, that are conditions imposed only at the
vertices of the graph:
\be
\sum_v\sum_{e_v}X^i_{e_v}\cdot{\cal
N}[\Upsilon_{{\bf j},{\bf l}} ,A] =0
\ee
where the first sum is
over vertices $v$ of the graph and the second over edges $e_v$ for
each vertex $v$. Spin networks that satisfy these conditions are
called gauge invariant spin networks (for a detailed treatment see
Sec.~4.1.5 of \cite{Perez:2004hj}). Finally, there is a very
convenient language for performing calculations involving (gauge
invariant) spin networks, that employs  graphical manipulations
over the vertices of the graphs. This is known as {\it recoupling
theory}, and (due to space restrictions here) it can be learned
from the primers by Major \cite{seth} and Rovelli
\cite{Rovelli:1998gg} and in Rovelli's Book \cite{Rovelli:2004tv}.

In next section we shall explore the picture of the quantum
geometry that the mathematical apparatus provides for us.

\section{Quantum Geometry}
\label{sec:5}

So far, we have constructed the kinematic Hilbert space of the
quantum theory. This is the quantum analog of the classical phase
space of the theory, the space of space-time metrics. Just as the
spacetime, together with its geometrical constructs such as
vectors, tensor fields, derivatives, etc. is the arena for doing
classical geometry, the space ${\cal H}_{\rm kin}$ is the arena
for quantum geometry. What we need is to define the quantum
object, i.e. the operators, that will correspond to the geometric
objects defined on this arena. Note that the background manifold
$\Sigma$ has many of the features of the classical geometry,
namely all the structure that comes with $\Sigma$ being a
differentiable manifold: tangent spaces, vectors, forms, etc. What
is {\it not} available are the notions arising from a metric on
$\Sigma$: distance, angles, area, volume, etc. These are the
properties of the geometry that are quantized. The purpose of this
section is to explore this quantum geometry.

This section has three parts. In the first one we describe the
nature of the operators associated to electric fluxes. In the
second part, we present the simplest of the geometric operators,
namely the area of surfaces and in the last part we discuss the
non-commutative character of the quantum geometry.

\subsection{Flux operators}

The operators $\hat{E}[S,f]$ corresponding to the electric flux
observables, are in a sense the basic building blocks for
constructing the quantum geometry. We have seen in Sec.\ref{sec:3}
the action of this operators on cylindrical functions,
\be
 \hat{E}[S,f]\cdot\Psi_\Upsilon(\overline{A})=-i\,\hbar\,
\{ \Psi_\Upsilon, \, {}^2\!E[S,f] \} = -i\,\frac{\ell^2_{\rm
P}}{2} \sum_{p}\, \sum_{I_p}\, \kappa({I_p})\, f^i(p)\,
X^i_{I_p}\cdot \psi\label{flux1} \ee
Here the first sum is over the intersections of the surface $S$
with the graph $\Upsilon$, and the second sum is over all possible
edges $I_p$ that have the vertex $v_p$ (in the intersection of $S$
and the graph) as initial of final point. In the simplest case of
a loop $\a$, there are only simple intersections (meaning that
there are two edges for each vertex), and in the simplest case of
only one intersection between $S$ and $\a$ we have one term in the
first sum and two terms in the second (due to the fact that the
loop $\a$ is seen as having a vertex at the intersection point).
In this simplified case we have
\be
 \hat{E}[S,f]\cdot\Psi_\a(\overline{A})=\,
-i\,\ell^2_{\rm P} \,   f^i(p)\, X^i_{I_p}\cdot \psi
 \ee
Note that the action of the operator is to `project' the angular
momentum in the direction given by $f^i$ (in the internal space
associated with the Lie algebra). As we shall see, this operator
is in a sense fundamental the fundamental entity for constructing
(gauge invariant) geometrical operators. For this, let us rewrite
the action of the flux operator (\ref{flux1}), dividing the edges
that are above the surface $S$, as `up' edges, and those that lie
under the surface as `down' edges. \be
\hat{E}[S,f]\cdot=\frac{\ell^2_P}{2}\sum_p
f^i(p)\,(\hat{J}^{p}_{i(u)}- \hat{J}^{p}_{i(d)})\cdot \ee where
the sum is over the vertices at the intersection of the graph and
the surface, and where the `up' operator
$\hat{J}^{p}_{i(u)}=\hat{J}^{p,e_1}_{i}+
\hat{J}^{p,e_2}_{i}+\cdots+\hat{J}^{p,e_u}_{i}$ is the sum over
all the up edges and the down operator  $\hat{J}^{p}_{i(d)}$ is
the corresponding one for the down edges.

\subsection{Area operator}

The simplest operator that can be constructed representing
geometrical quantities of interest is the {\it area operator},
associated to surfaces $S$. The reason behind this is again the
fact that the densitized triad is dual to a two form that is
naturally integrated along  a surface. The difference between the
classical expression for the area and the flux variable is the
fact that the area is a gauge invariant quantity. Let us first
recall what the classical expression for the area function is, and
then we will outline the regularization procedure to arrive at a
well defined operator on the Hilbert space. The area $A[S]$ of a
surface $S$ is given by $A[S]=\int_S\d^2 x\,\sqrt{h}$, where $h$
is the determinant of the induced metric $h_{ab}$ on $S$. When the
surface $S$ can be parametrized by setting, say, $x^3=0$, then the
expression for the area in terms of the densitized triad takes a
simple form: \be A[S]=\gamma\int_S\d^2
x\,\sqrt{\tilde{E}^3_i\,\tilde{E}^3_j\,k^{ij}} \ee where
$k^{ij}=\delta^{ij}$ is the Killing-Cartan metric on the Lie
algebra, and $\gamma$ is the Barbero-Immirzi parameter (recall
that the canonical conjugate to $A$ is
${}^\gamma\!\tilde{E}^a_i=\tilde{E}^a_i/\gamma$). Note that the
functions is again smeared in two dimensions and that the quantity
inside the square root is very much a square of the (local) flux.
One expects from the experience with the flux operator, that the
resulting operator will be a sum over the intersecting points $p$,
so one should focus the attention to the vertex operator
$$\Delta_{S,\Upsilon,p}=-\left[(\hat{J}^{p}_{i(u)}-
\hat{J}^{p}_{i(d)})(\hat{J}^{p}_{j(u)}-
\hat{J}^{p}_{j(d)})\right]k^{ij}$$
with this, the area operator takes the form,
\be
\hat{A}[S]=\gamma\,\ell^2_{\rm P}\sum_p
\widehat{\sqrt{\Delta_{S,\Upsilon,p}}}
\ee
We can now combine both the form of the vertex operator with Gauss' law
$(\hat{J}^{p}_{i(u)}+ \hat{J}^{p}_{i(d)})\cdot \Psi=0$ to arrive at,
\be
|(\hat{J}^{p}_{i(u)} - \hat{J}^{p}_{i(d)})|^2 = |2(\hat{J}^{p}_{i(u)})|^2
\ee
where we are assuming that there are no tangential edges. The
operator $\hat{J}^{p}_{i(u)}$ is an angular momentum operator, and
therefore its square has eigenvalues equal to $j^u(j^u+1)$ where
$j^u$ is the label for the total `up' angular momentum. We can
then write the form of the operator \be \hat{A}[S]\cdot{\cal
N}(\Upsilon,\vec{j})=\gamma\,\ell^2_{\rm P}
 \sum_{v\in \,V}
 \;\sqrt{|\hat{J}^{p}_{i(u)}|^2}\cdot
{\cal N}(\Upsilon,\vec{j})
\ee
With these conventions, in the case of  simple intersections
between the graph $\Upsilon$ and the surface $S$, the area
operator takes the well known form:
\[
\hat{A}[S]\cdot{\cal N}(\Upsilon,\vec{j})=\gamma\,\ell^2_{\rm P}
\sum_{v\in \,V}
 \;\sqrt{j_v(j_v+1)}\cdot
{\cal N}(\Upsilon,\vec{j})
\]
when acting on a {\it spin network} ${\cal N}(\gamma,\vec{j})$
defined over $\Upsilon$ and with labels $\vec{j}$ on the edges (we
have not used a label for the intertwiners).

Let us now interpret these results in view of the new geometry
that the loop quantization gives us. The one dimensional
excitations of the geometry carry flux of area: whenever the graph
pierces a surface  it endows $S$ with a quanta of area depending
on the value of $j$. Furthermore, the eigenvalues of the operator
are discrete, giving a precise meaning to the statement that the
geometry is quantized: there a minimum (non-zero) value for the
area given by taking $j=1/2$ in the previous formula. Thus the
area gap $a_{\rm o}$ is given by \be a_{\rm o}=\gamma\,\ell_{\rm
P}^2\frac{\sqrt{3}}{2} \ee
 If the value of $\gamma$  is of order
unit, then we see that the minimum area is of the order of the
Planck area. In order to get a macroscopic value for area we would
need a very large number of intersections. The Immirzi parameter
has to be fixed to select the physical sector of the theory. The
current viewpoint is that the black hole entropy calculation can
be used for that purpose.

There are operators corresponding to other geometrical objects
such as volume, length, angles, etc. The main feature that they
exhibit is that their spectrum is always discrete.

\subsection{Non-commutative quantum geometry}

One of the somewhat unexpected results coming from loop quantum
geometry is the fact that the quantum geometry is inherently
non-commutative. This means that the  operators associated to
geometrical quantities do not, in general, commute.

Let us for simplicity consider area operators. Given a surface
$S$, any spin network state defined on graphs that do not have
vertices on the surface are eigenstates of the area operator
(bi-valent vertices are equivalent to no-vertices). There is no
degeneracy. If the graph has trivalent vertices on the surface,
spin networks are still eigenstates and there is again no
degeneracy. It is for a four valent vertex that things start to
get interesting. It is {\it not} true that any spin network state
is an area eigenstate. There is, however, a choice of basis of
eigenstates for any given area operator. This is true for any
valence, namely that given one surface and any vertex thereon
there is always a choice of basis that is an eigenbasis for that
surface. There will be, however, degeneracy for higher valence
vertices. For a four valent one, we know that for a fixed coloring
of the edges, there will be some degeneracy. That is, the (finite
dimensional) vector space spanned by the intertwiners can have
dimension grater that one. The dimension will be given by the
number of possible representations of $SU(2)$ compatible with the
external colorings (via the triangle inequality). In order to
break the degeneracy, it is enough to have one surface operator
$A[S]$ acting on the vertex, provided that it has two vertices on
each side of the surface. That choice of surface will then
`select' the decomposition of the vector space into eigenstates of
$A[S]$, the eigenvalues given by the $\sqrt{j_{\rm int}(j_{\rm
int}+1)}$ kind of formula, where the $j_{\rm int}$ means the
`internal' coloring of the representation used to write the
intertwiner.

Let us now consider an explicit case of non-commutativity for a
graph $\Upsilon$ having a four valent vertex $v$ at the surface
$S$. We assume that the edges $e_1$ and $e_2$ are {\it up} edges
for $S_1$ (and $(e_3,e_4)$ are down). We denote by $j_5$  the
internal edge, and we choose as the state a linear combination of
spin networks (with coefficients $(\a_0,\a_1,\a_2)$) of spin
networks having $j_5=0,j_5=1$ and $j_5=2$ respectively as the
internal index: ${\Psi}(\Upsilon,{\bf j_0})=\a_{0}\;{\cal
N}(\gamma,{j}_5=0)+\a_{1}\;{\cal N}(\gamma,{j}_5=1)+\a_2\,{\cal
N}(\gamma,{j}_5=2)=\sum_i\a_i\,{\cal N}_i$. The external edges
$(j_1,j_2,j_3,j_4)$ are all taken to be equal to one. For
simplicity let us denote by ${\cal N}_i$ the eigenstates of the
area operator: $\hat{A}[S_1]\cdot {\cal N} _i=a_i\,{\cal N}_i$,
where $a_0=0,a_1=\gamma\ell^2_{\rm P}\,\sqrt{2}$ and
$a_3=\gamma\ell^2_{\rm P}\,\sqrt{6}$.

First we need to compute the action of the first surface
$\hat{A}[S_1]$,
\be \hat{A}[S_1]\cdot {\Psi}(\Upsilon,{\bf
j_0})=\sum_i\a_i\,a_i\,{\cal N}_i \ee We know that in this case
the vector space of intertwiners is three dimensional (spanned by
${\cal N}_i$), so we can think of the area operator acting on this
space as a matrix $A^i_j=a_i\,\delta^i_j$ (no summation over $i$).

Before acting with $\hat{A}[S_2]$ we need to change basis from the
basis diagonal on $j_5$ to the one diagonal to $j_6$. Thus we need
to expand ${\cal N}_i$ in terms of the other basis. The $3 \times
3$ matrix ${U^j}_k$ that changes basis has to be an element of
$SO(3,\mathbb{C})$. Its precise form is not relevant in what
follows but it is known that its components are given by $6-j$
symbols \cite{Rovelli:2004tv}. We can then write, \be {\cal
N}_j={U^k}_j\,\tilde{\cal N}_k \ee We can now compute
 \ba
 \hat{A}[S_2]\cdot \hat{A}[S_1]\cdot
{\Psi}(\Upsilon,{\bf j_0}) &=&  \hat{A}[S_2]\cdot\sum_{i,k}
\a_i\,a_i {U^k}_i\,\tilde{\cal N}_k \\
&=& \sum_{i,k} \a_i\,a_i {U^k}_i\,a_k\,\tilde{\cal N}_k \ea
Let us now act with the operators in the reverse order, namely let
us act with $\hat{A}(S_2)$ first,
 \be
 \hat{A}[S_2]\cdot{\Psi}(\Upsilon,{\bf j_0})=\sum_{i,k}\a_i{U^k}_i\,a_k
 \,\tilde{\cal N}_k
\ee Now, before acting with $\hat{A}[S_1]$ we need to change basis
from the basis diagonal on $j_6$ to the basis diagonal to $j_5$.
Thus we need to expand $\tilde{\cal N}_i$ in terms of the other
basis. The matrix that defines this change of basis will be the
inverse ${U^{(-1)k}}_j$ such that $\tilde{\cal
N}_j={U^{(-1)k}}_j\,{\cal N}_k $. We can now compute
 \ba
 \hat{A}[S_1]\cdot \hat{A}[S_2]\cdot
{\Psi}(\Upsilon,{\bf j_0}) &=&  \hat{A}[S_1]\cdot\sum_{i,k,m}
\a_i\, {U^k}_i\,a_i  {U^{(-1)m}}_k\,{\cal N}_m  \\
&=& \sum_{i,k,m} \a_i\, {U^k}_i\,a_i{U^{(-1)m}}_k\,a_m\,{\cal N}_m
\ea We can then write the commutator as,
\be [\hat{A}[S_1], \hat{A}[S_2]]\cdot{\Psi}(\Upsilon,{\bf j_0}) =
\sum_{i,k,m} \a_i\, \left({U^k}_i\,a_i{U^{(-1)m}}_k\,a_m- a_i\,
{U^k}_i\,a_k\,{U^{(-1)m}}_k\,\right)\,{\cal N}_m \ee
Which is, in general, different from zero (for instance, the area
matrix $A^{i}_{j}=a_i\delta^i_j$ does not commute with ${U^i}_k$).
This implies clearly that there is an uncertainty relation for
intersecting areas
$$(\Delta \hat{A}[S_1])^2\,(\Delta \hat{A}[S_2])^2\geq 0$$
For more details on non-commutativity see \cite{non-comm}.

\section{Results and open issues}
\label{sec:6}

So far we have dealt with the kinematical aspects of the theory,
in which the states are (gauge invariant) states labelled by
embedded graphs on the three manifold $\Sigma$. However, the fact
that general relativity is diffeomorphism invariant as a classical
field theory provides a good motivation for trying to implement
diffeomorphism invariance at the quantum level. In the $3+1$
setting this means implementing the constraints as quantum
conditions on the quantum states.

Let us first consider the diffeomorphism constraint, that is, the
generator of spatial diffeomorphisms. In the canonical setting,
this means that we should ask for diffeomorphism invariant states.
The answer is rather easy: the only diffeomorphism invariant state
in ${\cal H}_{\rm kin}$ is the vacuum $\Psi_0$! This means that
the intuitive idea that diff-invariant states as obtained by
taking the quotient by the diffeomorphisms in the Hilbert space
does not work. One needs to find a subtler prescription.
Fortunately, it is now well understood that, in general, solutions
to the quantum constraints do not lie on the original kinematical
Hilbert space where the constraints were originally posed.
Instead, one has to look for solutions in a larger space. In the
case of LQG, diffeo-invariant states live  in $({\rm Cyl})^*$, the
dual space to the cylindrical functions, where a solution is found
by summing over all possible diffeomorphism related states (recall
that the action of the diffeomorphisms is a unitary map in the
kinematical Hilbert space). Even when the sum is over an
uncountable number of states, as a {\it distribution} its action
on cylindrical functions is well defined. The procedure also
includes specifying a diffeo-invariant inner product on solutions,
and by completing, one arrives at ${\cal H}_{\rm diff}$, the
Hilbert space of diffeomorphism invariant states. Roughly
speaking, instead of considering states labelled by embedded
graphs as in ${\cal H}_{\rm kin}$, diffeo-invariant states are
labelled by diffeomorphism classes of graphs (apart from the
internal labellings $(\bf{j},\bf{i})$). Having solved the
diffeomorphism invariance, in a rigorous manner and without
anomalies, is one of the main achievements of the formalism. For
details see \cite{AL:review}.

Due to lack of space, we will refrain from discussing in detail
two of the main applications of LQG, namely Black Holes and
Cosmology, so we will only give a brief summary. For black holes,
a detailed treatment of the boundary conditions of the theory in
the presence of an isolated horizon has lead to a detailed
description of the quantum geometry associated with the horizon,
to an identification of the quantum degrees of freedom responsible
for the entropy, and to a counting of states that yields the
entropy proportional to the area. The proportionality factor is
set, for large black holes, to $1/4$ by adjusting the only free
parameter of the theory, that is, the Barbero-Immirzi parameter.
The same value can account for the entropy of very general BH,
including Einstein-Maxwell, Einstein-Dilaton and non-minimally
coupled scalar fields. For details and references see
\cite{AL:review}.

In the case of cosmology, a detailed program known as {\it Loop
Quantum Cosmology} (LQC), has been developed in the past years
(see for instance Bojowald's contribution to this volume and
\cite{bojo}). It represents a dimensional reduction, at the
classical level, of the gravitational degrees of freedom to
instances of great symmetry, as is expected to be the case in the
cosmological setting. The quantization is performed via
non-standard representations of the classical observable algebra,
much in the spirit of LQG, with certain input coming from the full
theory. As a result, the theory possesses very different behavior
as the standard mini-superspace quantization: the space-time
curvature has been shown to be bounded above and thus, the
classical singularity is avoided. Furthermore, there is some
evidence that the solutions may exhibit an inflationary period.

More recently, a new physical picture for the evaporation of black
holes has emerged both from a detailed treatment of dynamical
horizons and the end result of gravitational collapse (borrowing
from results in LQC). These new developments suggests that one
should reconsider the traditional viewpoint towards Hawking
evaporation and the information loss problem \cite{AA:MB}.


Loop quantum gravity is a theory in the making. There are still
unsolved issues and things that remain to be done. The most
notorious issue that has eluded a complete solution is the
implementation of the Hamiltonian constraint. There is no proposal
that is fully satisfactory. There is a consistent, free of
anomalies quantization by Thiemann, but is far from unique and
there is no control (yet) on the physical properties of its
solutions (see \cite{Thiemann:2001yy}). More recently Thiemann has
proposed the `master constraint program' to implement the
constraints. There is a completely different approach to the
subject, where a projector onto physical states is sought by
summing over 4-dim discrete structures known as spin-foams (see
for instance \cite{Perez:2004hj}). Finally, a new program that
tries to solve the constraint in a similar fashion as the
diffeomorphism invariance is implemented, namely by considering
finite actions of the constraint rather than defining and
implementing its generator, is being pursued nowadays.

Another open issue is the semi-classical limit of the theory. That
is, one needs to find states within the theory that approximate,
in a low-energy/macroscopic limit, classical configurations such
as Minkowski spacetime. To deal with this challenge there have
been several proposals for semi-classical states, from the
so-called weave states, to coherent states \cite{Thiemann:2001yy},
WKB-like states \cite{Smolin:2004sx}, and more recently
Fock-motivated states \cite{AL:review}. In most of these cases,
the states belong to the kinematical space of states. A systematic
approach to the definition of semiclassical states for constrained
systems and to the question of whether one can approximate
physical semi-classical states with kinematical ones has only
begun \cite{ABC}.

In previous sections we have defined LQG in terms of what it is
and does. To end this section we will part from this tradition and
make some remarks about what LQG is {\it not}. This can be seen as
a complement to the FAQ section of \cite{Smolin:2004sx}.

\begin{itemize}

\item {\it Renormalizability}.
 It is sometimes asserted that since LQG is a reformulation of
General Relativity, then it is non-renormalizable. The first
obvious observation is that LQG {\it is} a non-perturbative
quantum theory, and as such, it has been shown to be finite (see
for instance \cite{AL:review,Thiemann:2001yy}). There is no
inconsistency with the fact that Einstein gravity is
perturbatively non-renormalizable. The level of rigor with which
LQG is defined, as a mathematically well defined theory, is
superior to any interacting quantum field theory in 4 dimensions.
As such, the theory has also shown to be quite independent of
``infrared" cutoffs.

\item {\it What is LQG not?} One sometimes sees statements like:
``various alternative formulations of LQG such as random lattice
or spin foam theory..." This claim is incorrect. Spin foams and
random lattices are not alternative formulations of LQG. In
particular, random lattices are not directly related to LQG, and
spin foams are tools (under development) to compute projection
operators onto physical states of the theory, i.e. as a tool for
solving the quantum scalar constraint of LQG (see
\cite{Perez:2004hj} for details).

\item {\it Is LQG a discrete theory?} In this regard, it is
important to stress, again, that LQG is a well defined theory on
the continuum. It is {\it not} a discretized gravity model such as
Regge calculus or dynamical triangulations, where the discrete
structure is assumed from the beginning. It is true that a
convenient basis of states in the Hilbert space of LQG is given by
graphs and spins (the spin networks basis), but the theory is
defined in the continuum, and the quantized character of the
geometry (the discrete eigenvalues for the operators) are found as
predictions of the theory. There is no continuum limit to be
taken. All details and mathematical proofs of these assertions can
be found in Refs.~\cite{AL:review,Thiemann:2001yy}.

\end{itemize}

To end this primer, let us point out to further references. Once
the reader is somewhat familiar with the basic assumptions of the
theory, she should move on to move detailed description of the
formalism. In this regard, the book by Rovelli
\cite{Rovelli:2004tv} is a good read for the conceptual background
and spirit of LQG. For a good degree of detail, rigor and current
status of the program see \cite{AL:review}. For a completely
detailed treatment of the formalism see the monograph by Thiemann
\cite{Thiemann:2001yy}.


\clearpage
\section*{Acknowledgments}

\noindent I would like to thank the organizers of the VI Mexican
School for the invitation  to write this contribution. This work
was in part supported by grants DGAPA-UNAM IN-108103 and CONACyT
36581-E.


\begin{thebibliography}{99}

\bibitem{pullin}
  Pullin J 1994 Knot theory and quantum gravity in loop Space: A Primer
  {\it AIP Conf.\ Proc.}  {\bf 317} 141
  ({\it Preprint} hep-th/9301028).


\bibitem{Rovelli:1998gg}
  Rovelli C and Upadhya P 1998
  Loop quantum gravity and quanta of space: A primer
  ({\it Preprint} gr-qc/9806079).

\bibitem{seth} Major S A 1999
  A spin network primer
  {\it Am.\ J.\ Phys.}  {\bf 67} 972
  ({\it Preprint} gr-qc/9905020)



\bibitem{AA:NJP} Ashtekar A 2004 Gravity and the quantum
  {\it Preprint} gr-qc/0410054

\bibitem{Smolin:2004sx}
  Smolin L 2004 An invitation to loop quantum gravity
  {\it Preprint} hep-th/0408048

\bibitem{AL:review} Ashtekar A and Lewandowski J 2004
  Background independent quantum gravity: A status report
  {\it Class.\ Quant.\ Grav.}  {\bf 21} R53
  ({\it Preprint} gr-qc/0404018)

\bibitem{Perez:2004hj}
  Perez A 2004
  Introduction to loop quantum gravity and spin foams
  {\it Preprint} gr-qc/0409061

\bibitem{Thiemann:2002nj}
  Thiemann T 2003 Lectures on loop quantum gravity
  {\it Lect.\ Notes Phys.}  {\bf 631} 41
  ({\it Preprint} gr-qc/0210094)

\bibitem{Nicolai:2005mc}
  Nicolai H, Peeters K and Zamaklar M 2005
  Loop quantum gravity: An outside view
 {\it Preprint} hep-th/0501114

\bibitem{Rovelli:2004tv}  Rovelli C 2004 {\it Quantum gravity}
(Cambridge, UK: Cambridge University Press)

\bibitem{Thiemann:2001yy} Thiemann T 2001 Introduction to modern canonical
quantum general relativity {\it Preprint} gr-qc/0110034

\bibitem{Rovelli:1999hz}
  Rovelli C 2000 The century of the incomplete revolution:
  Searching for general  relativistic quantum field theory
  {\it J.\ Math.\ Phys.}  {\bf 41} 3776
  ({\it Preprint} hep-th/9910131)

\bibitem{Rovelli:1997yv}  Rovelli C 1998
  Loop quantum gravity {\it Living Rev.\ Rel.}  {\bf 1} 1
  ({\it Preprint} gr-qc/9710008)


\bibitem{pullin2} Pullin J 2003
  Canonical quantization of general relativity: The last 18 years in a
  nutshell {\it AIP Conf.\ Proc.}  {\bf 668} 141 (2003)
  ({\it Preprint} gr-qc/0209008)

\bibitem{Rovelli:1997qj}
Rovelli C 1998 Strings, loops and others: A critical survey
  of the present approaches to quantum gravity
  {\it Preprint} gr-qc/9803024

\bibitem{Rovelli:2000aw}
  Rovelli C 2000
  Notes for a brief history of quantum gravity
  {\it Preprint} gr-qc/0006061

\bibitem{Rovelli:2003wd}
  Rovelli C 2003  A dialog on quantum gravity
  {\it Int.\ J.\ Mod.\ Phys.\ D} {\bf 12} 1509
  ({\it Preprint} hep-th/0310077)

\bibitem{Smolin:2003rk}
  Smolin S 2003 How far are we from the quantum theory of gravity?
  {\tt Preprint} hep-th/0303185

\bibitem{seth2}
{\tt
http://academics.hamilton.edu/physics/smajor/Papers/read\_guide.html}

\bibitem{biblio} See for instance, Beetle C and Corichi A 1997
Bibliography of publications related to classical and quantum
gravity in terms of connections and loop variables {\it Preprint}
gr-qc/9703044
 For an updated version see {\tt
http://www.nucleares.unam.mx/$\;\tilde{ }$corichi/lqgbib.pdf}

\bibitem{wald1} Wald R M 1984 {\it General Relativity}
(Chicago: Chicago University Press)

\bibitem{poisson} Poisson E 2003 {\it A Relativist Toolkit} (Cambdridge,
UK: Cambridge U. Press)

\bibitem{wald2} Wald R M 1994
{\it Quantum field theory in curved space-time and black hole
thermodynamics} (Chicago: Chicago University Press)

\bibitem{baez} Baez J and Muniain J P 1994
{\it Gauge fields, knots and gravity}
(Singapore: World Scientific)

\bibitem{barbero} Barbero J F 1995
  Real Ashtekar variables for Lorentzian signature space times
  {\it Phys.\ Rev.\ D} {\bf 51} 5507
  ({\it Preprint} gr-qc/9410014)

\bibitem{Immirzi} Immirzi G 1997 Quantum gravity and Regge calculus
{\it Nucl.\ Phys.\ Proc.\ Suppl.}  {\bf 57} 65
({\it Preprint} gr-qc/9701052)

\bibitem{ACZ}
Ashtekar A, Corichi A and Zapata J A 1998
Quantum theory of geometry. III: Non-commutativity of Riemannian
  structures {\it Class.\ Quant.\ Grav.}  {\bf 15} 2955
  ({\it Preprint} gr-qc/9806041)

\bibitem{CCQ} Corichi A, Cortez J and Quevedo H 2004 On the relation
 between Fock and Schr\"odinger representations for a  scalar field
 {\it Annals Phys.}  {\bf 313} 446  ({\it Preprint} hep-th/0202070)



\bibitem{Velhinho}
  Velhinho J M 2004 On the structure of the space of generalized
  connections
  {\it Int.\ J.\ Geom.\ Meth.\ Mod.\ Phys.}  {\bf 1} 311
  ({\it Preprint} math-ph/0402060)

\bibitem{LOST} Lewandowski J, Okolow A, Sahlmann H and Thiemann T
2005 Uniqueness of diffeomorphism invariant states on
holonomy-flux algebras
 {\it Preprint}  gr-qc/0504147

\bibitem{poly} Ashtekar A Fairhurst S and Willis J L 2003
 Quantum gravity, shadow states, and quantum mechanics
  Class.\ Quant.\ Grav.\  {\bf 20} 1031
  ({\it Preprint}  gr-qc/0207106).


\bibitem{LQC} Ashtekar A, Bojowald M and Lewandowski J 2003
  Mathematical structure of loop quantum cosmology
  {\it Adv.\ Theor.\ Math.\ Phys.}  {\bf 7} 233
  ({\it Preprint} gr-qc/0304074)

\bibitem{bojo} Bojowald M 2005
  Elements of loop quantum cosmology {\it Preprint}
  gr-qc/0505057


\bibitem{alm2t} Ashtekar A, Lewandowski J, Marolf D, Mourao J
and Thiemann T 1995 Quantization of diffeomorphism invariant theories
of connections with local degrees of freedom
  {\it J.\ Math.\ Phys.}  {\bf 36} 6456
  ({\it Preprint} gr-qc/9504018)

\bibitem{baez2} Baez J C 1996 Spin network states in gauge theory
  {\it Adv.\ Math.}  {\bf 117} 253  ({\it Preprint} gr-qc/9411007)

\bibitem{non-comm} Corichi A and Zapata J A 2005  On the non-commutative
geometry of loop quantum gravity {\it Preprint}

\bibitem{AA:MB} Ashtekar A and Bojowald M 2005 Black hole evaporation:
A paradigm {\it Preprint} gr-qc/0504029

\bibitem{ABC} Ashtekar A, Bombelli L and Corichi A 2005
Semiclassical states for constrained systems {\it Preprint}
gr-qc/0504052

\end{thebibliography}
\end{document}